\documentclass[sn-mathphys,Numbered]{sn-jnl}

\usepackage{graphicx}%
\usepackage{multirow}%
\usepackage{amsmath,amssymb,amsfonts}%
\usepackage{amsthm}%
\usepackage{mathrsfs}%
\usepackage[title]{appendix}%
\usepackage{xcolor}%
\usepackage{textcomp}%
\usepackage{manyfoot}%
\usepackage{booktabs}%
\usepackage{algorithm}%
\usepackage{algorithmicx}%
\usepackage{algpseudocode}%
\usepackage{listings}%

\theoremstyle{thmstyleone}%
\theoremstyle{thmstyletwo}%

\theoremstyle{thmstylethree}%

\begin{document}

\title{Dispersion Managed Elliptical Atomtronics for Interferometry}


\author[1]{\fnm{Sriganapathy} \sur{Raghav}}

\author[2]{\fnm{Suranjana} \sur{Ghosh}}

\author[3]{\fnm{Jayanta} \sur{Bera}}

\author*[1]{\fnm{Utpal} \sur{Roy}}\email{uroy@iitp.ac.in}

\affil*[1]{\orgdiv{Department of Physics}, \orgname{Indian Institute of Technology}, \orgaddress{ \city{Patna}, \postcode{801106}, \state{Bihar}, \country{India}}}

\affil[2]{\orgdiv{Department of Physics}, \orgname{Indian Institute of Science Education and Research}, \orgaddress{\city{Kolkata}, \postcode{741246}, \state{West Bengal}, \country{India}}}

\affil[3]{\orgdiv{Department of Physics}, \orgname{C. V. Raman Global University}, \orgaddress{\city{Bhubaneshwar}, \postcode{752054}, \state{Odisha}, \country{India}}}


\abstract{
Circular atomtronics is known to exhibit a uniform ground state, unlike elliptical atomtronics. In elliptical atomtronics, the matter wave tends to accumulate along the semi-major edges during its time dynamics, which we depict by the survival function. Consequently, the dynamical time scales become coupled to the eccentricity, making the dynamics nontrivial for applications. We report that, an appropriate dispersion management can decouple the time scales from the eccentricity. One can choose the suitable dispersion coefficient from the overlap function involving the corresponding ground state. We focus on producing distinct fractional matter waves inside an elliptical waveguide to achieve efficient atom-interferometry. The said dispersion engineering can recover fractional revivals in the elliptical waveguide, analogous to the circular case. We demonstrate atom-interferometry for the engineered elliptical atomtronics, where matter wave interference is mediated by an external harmonic trap for controlled interference patterns.}

\maketitle

\section{Introduction}

Guiding matter waves is a pivotal component in constructing matter wave circuits, particularly within atom chip technology \cite{amico2021roadmap,amico2022colloquium,amico2017focus}. The precise fabrication of these waveguides is critical in the field of atomtronics, allowing coherent control and manipulation of matter waves \cite{ryu2015integrated,pandey2021atomtronic}. Techniques like time-averaged adiabatic potential (TAAP) \cite{lesanovsky2007time,gildemeister2010trapping,henderson2009experimental,pandey2019hypersonic,navez2016matter}, intensity mask \cite{lee2014spatial}, and digital holography \cite{gaunt2012robust} have been instrumental in creating waveguides for ultracold atoms, contributing significantly to the advancement of atomtronics. Among the various types of waveguides, the circular waveguide stands out as the simplest spatially closed atomtronic circuit \cite{gupta2005bose,bell2016bose}, widely applied in atom interferometry \cite{berman1997atom,baudon1999atomic,fattori2008atom,petrovic2013multi,sewell2010atom}, quantum transport \cite{dutta2006single,haug2019andreev}, quantum sensing \cite{ragole2016interacting,pelegri2018quantum,kialka2020orbital}, atom SQUID \cite{ryu2013experimental,mathey2016realizing}, and other quantum technological applications \cite{amico2022colloquium}. The curvature-induced potential (CIP) of a waveguide is proportional to the square of its curvature, implying a constant CIP for a circular waveguide \cite{schwartz2006one,sandin2017dimensional,tononi2023low}, which bears a uniformly distributed ground state along its circumference \cite{salasnich2022bose}.

The major utility of a circular waveguide stems from its constant curvature, which facilitates matter wave interference, an integral component of the above applications. Interference of matter waves in a circular waveguide also leads to fractional revivals (FR), which is the spawning of multiple replicas of the initial matter wave packet along the perimeter of the circular waveguide \cite{averbukh1989fractional,bera2020matter,raghav2023matter}. The FR instances are succeeded by the revival of the initial condensate in shape and location. The FR time instances and revival depend on a characteristic parameter linked to the radius of the waveguide \cite{bera2020matter}. Moreover, higher-order FR patterns can provide a platform for studying multiple source interference, which is usually studied using optical lattices \cite{pedri2001expansion,zhang2002interference,hadzibabic2004interference,ashhab2005interference}. In contrast, due to the variable width along the circumference or non-constant CIP, elliptical waveguides result in a nonuniform ground state along the perimeter \cite{salasnich2022bose,Nikolaieva_2023}. Unlike a circular waveguide, the elliptical counterpart lacks support for Talbot oscillations \cite{campo2014bent}.

Concurrently, dispersion engineering has emerged as a significant method for controlling and manipulating the dynamics of Bose-Einstein condensates (BEC) in external traps \cite{eiermann2003dispersion,su2022self,das2016realization}. One can control the dynamics inside a matter-wave circuit by tuning the nonlinearity and the dispersion through Feshbach resonance and DM, as in the case of optical solitons \cite{malomed2006soliton,mayteevarunyoo2020spatiotemporal}. Recently, the nonlinearity of the BEC was used to nullify the effects of curvature in an elliptical waveguide \cite{Nikolaieva_2023}. Here, we delve into engineering the dispersion of BEC inside an elliptical waveguide to regain the properties of circular atomtronics. Notably, the utilization of matter wave dispersion serves as a means to counteract the effects of non-constant width in an elliptical waveguide, offering potential strategies for manipulating and optimizing matter wave behaviour in atomtronics applications.

The paper is organized as follows: Section II introduces the theoretical model for BEC in an elliptical waveguide. This section also discusses the role of waveguide geometry in the dispersion of matter waves. In Sec. III, we derive the dynamical equation for BEC with tunable dispersion in an elliptical waveguide and also outline the numerical methods applied to obtain both the ground state solution and FR dynamics within the waveguide. The ground state solution of BEC in an elliptical waveguide, characterized by different eccentricities, are demonstrated. We also determine the suitable dispersion coefficients for matter waves, aiming to neutralize the effects of variable thickness in an elliptical waveguide. Section IV delves into exploring the impact of eccentricity on the FR dynamics of matter waves and the subsequent restoration of FR signatures through dispersion management, which elucidates how manipulating dispersion can counteract the challenges posed by the non-constant width in elliptical waveguides. In Section V, we focus on the atom interferometry for dispersion managed elliptical waveguide. Finally, we conclude in Sec. VI, summarizing the key findings and future outlook.

\section{Bose-Einstein Condensate in an Elliptical Waveguide}

We consider a BEC of $N$ number of ${}^{23}$Na atoms, loaded in an elliptical waveguide of eccentricity $\varepsilon$. A three-dimensional Gross Pitaevskii equation (3D-GPE) \cite{pethick2008bose,pitaevskii2016bose,dalfovo1999theory,adhikari2009localization,atre2006class} describes the evolution of the macroscopic wavefunction $\Psi\equiv \Psi(r,t)$ of a BEC in an external trap:
\begin{eqnarray}
	i\hbar\frac{\partial \Psi}{\partial t}=\Bigg[-\frac{\hbar^2}{2m}\nabla^2+\frac{4\pi Na_{s}}{m}|\Psi|^2+V(r)\Bigg]\Psi, \label{3DGPE}
\end{eqnarray}
where $V(r)=V(z)+V(x,y)$ with $V(z)=\frac{1}{2}m\omega_{\perp} z^2\label{harmonic}$ and
\begin{eqnarray}
	V(x,y)=V_{0}\Bigg\{1-\exp{\Bigg[-\frac{1}{\gamma^2}\Bigg(\sqrt{x^2+\frac{y^2}{1-\varepsilon^2}}-a\Bigg)^2\Bigg]}\Bigg\}.\label{potential}\nonumber
\end{eqnarray}
Here, $\omega_{\perp}$ is the frequency of the transverse harmonic trap, $V_0$ is the depth of the waveguide, $\gamma$ is the width of the waveguide, and $\varepsilon=\sqrt{1-\frac{b^2}{a^2}}$ is the eccentricity. For a circular waveguide with a radius $a$, we take $\varepsilon=0$, and for an elliptical waveguide with a semi-major radius $a$, we consider a non-zero eccentricity.

The 3D GPE in Eq.\ref{3DGPE} is reduced to effective 2D GPE by considering a strong trap in the transverse direction by writing the wavefunction as $\Psi(r,t)=\psi(x,y,t)\Phi(z)$. Here, the function $\psi(x,y,t)$ describes the dynamics of BEC in the elliptical waveguide and $\Phi(z)$ denotes the ground state of the strong axial trap, which is a Gaussian wavefunction with width $a_{\perp}$ \cite{adhikari2019vortex}. The effective 2D GPE in the dimensionless form is obtained after integrating out the z component and by scaling position, time and energy by $a_{\perp}$, $\frac{1}{\omega_{\perp}}$ and $\hbar \omega_{\perp}$, respectively. Here $a_{\perp}=\sqrt{\frac{\hbar}{m \omega_{\perp}}}$ and $\omega_{\perp}$ are the harmonic oscillator length and frequency in the transverse direction, respectively.
\begin{equation}
	i\frac{\partial \psi}{\partial t}=\Bigg[-\frac{1}{2}\nabla_{x,y}^2+g|\psi|^2+V(x,y)\Bigg]\psi\label{2DGPE}
\end{equation}
Here, $g=\frac{2\sqrt{\pi}Na_{s}}{a_{\perp}}$ and, $V(x,y)$ is the potential of the elliptical waveguide scaled by $\hbar \omega_{\perp}$. This elliptical waveguide has varying thickness for non-zero eccentricity and manifests nontrivial dynamics.

\begin{figure}[ht]
	\centering
	\includegraphics[width=10 cm]{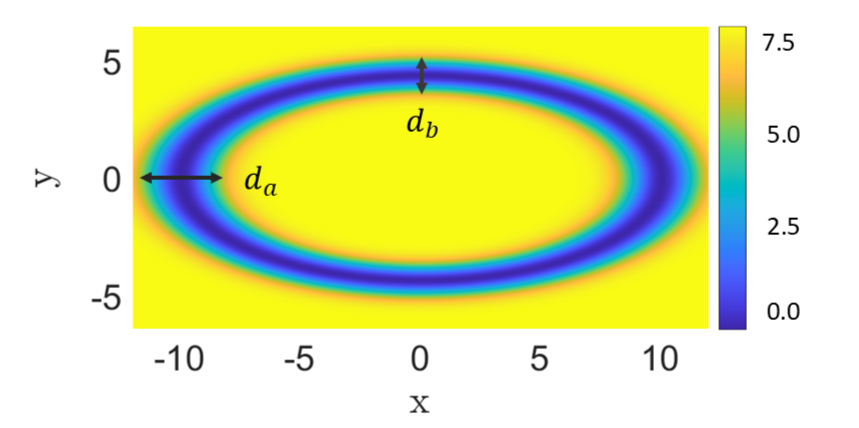}
	\caption{Elliptical ring trap with varying width along its circumference. The semimajor radius is taken as $a=10a_{\perp}$, and the semiminor radius is $b=4.36a_{\perp}$. The corresponding eccentricity is $\varepsilon=0.9$. $x$ and $y$ are in the units of $a_{\perp}=2.32\;\mu$m and, $t$ is in the units of $1/\omega_{\perp}=1.95\;$ms. The colorbar gives the value of potential in the units of $\hbar \omega_{\perp}$.}
	\label{Waveguide}
\end{figure}

\subsection{Waveguide Geometry and Matter Wave Dispersion}

A circular waveguide with $\varepsilon=0$, formed by a ring Gaussian potential, has constant width \cite{murray2013probing} unlike an elliptical waveguide, as shown in Fig.\ref{Waveguide}. This varying width of the ring affects the dispersion across different waveguide segments. To elucidate the influence of nonzero eccentricity on dispersion, we study the dynamics of a Gaussian wavepacket placed at two different positions inside the elliptical waveguide, namely $(a,0)$ and $(0,b)$. For the sake of simplicity, we consider the dynamics of a non-interacting cloud. The wavepacket is expressed as
\begin{equation}
	\chi(s,0)=\Bigg(\frac{d}{\sqrt{\pi}}\Bigg)^{\frac{1}{2}}\int_{-\infty}^{\infty} \frac{dk}{\sqrt{2\pi}} e^{-\frac{d^2k^2}{2}} e^{iks}
\end{equation}
Here $s$ is the coordinate along the axis of propogation of the wavepacket. By decomposing the initial wavefunction into its constituent Fourier modes, we write
\begin{eqnarray}
	\chi(s,t)= \Bigg(\frac{d}{\sqrt{\pi}}\Bigg)^{\frac{1}{2}}\int_{-\infty}^{\infty} \frac{dk}{\sqrt{2\pi}} e^{-\frac{d^2k^2}{2}} e^{-i\frac{E_{k}t}{\hbar}} e^{iks},\label{2}
\end{eqnarray}
where, $d$ is related to the width of the wavepacket, and $k$ relates the energy: $E_{k}=\hbar^2 k^2/(2m)$, which allows us to simplify the wavefunction in Eq.\ref{2} as,
\begin{eqnarray}
	\chi(s,t)=\Bigg(\frac{D}{\sqrt{\pi} d}\Bigg)^{\frac{1}{2}}e^{-\frac{Ds^2}{2d^2}}
\end{eqnarray}
with $D=\frac{1}{(1+i\frac{\hbar t}{md^2})}$. Consequently, the width at time $t$ becomes
\begin{eqnarray}
	w_{t}^2=\int_{-\infty}^{\infty}\chi^{*}(s,t)s^2\chi(s,t)ds=w_{0}^2+\frac{\hbar^2t^2}{4m^2w_{0}^2}.\label{widthev}
\end{eqnarray}
Here, the initial width is given by $w_{0}=\frac{d}{\sqrt{2}}$ and the above expression reduces to a dimensionless form as
\begin{equation}
	\tilde{w}_{t}^2=\tilde{w}_{0}^2+\frac{\tilde{t}^2}{4\tilde{w}_{0}^2}. \label{widthevol}
\end{equation}
Here $\tilde{w}_{t}=\frac{w_{t}}{a_{\perp}}$, $\tilde{w}_{0}=\frac{w_{0}}{a_{\perp}}$ and $\tilde{t}=t\omega_{\perp}$. For simplicity, we will omit the tildes from Eq.(\ref{widthevol}) onward, assuming all quantities are dimensionless.\\
For analysing the dispersions corresponding to the initial positions, $(a,0)$ and $(0,b)$, along the elliptical waveguide, we write the respective widths of the waveguide as $d_a$ and $d_b$, having ratio $\frac{d_b}{d_a}=\sqrt{1-\varepsilon^2}$. An intuitive understanding and numerical trials suggest we take the widths of the wavepacket perpendicular to the circumference of the waveguide as $d_a$ and $d_b$, respectively, to prevent inhomogeneous dispersion.

Accordingly, the normalization of the wavepacket at these places dictates that their widths along the circumference of the waveguide, $w_{a}$ and $w_{b}$, should be inversely related to $d_a$ and $d_b$:
\begin{equation}
	\frac{w_a}{w_b}=\frac{d_b}{d_a}=\sqrt{1-\varepsilon^2},\label{widths}
\end{equation}
which facilitates us to write their relation at time $t$, following Eq.\ref{widthevol} as
\begin{eqnarray}
	w_{b,t}^2=\frac{1}{1-\varepsilon^2}\Bigg[w_{a}^2+\frac{t^2(1-\varepsilon^2)^2}{4w_{a}^2}\Bigg].\label{corrwidthb}
\end{eqnarray}

It becomes apparent that, the widths of the wavepacket at the semi-major and semi-minor edges deviate from their initial ratio (Eq.\ref{widths}). The factor $(1-\varepsilon^2)^2$ introduces the inhomogeneity of the widths over the time. This demands a modification of $E_k$ by the same factor, such that $E_k=\frac{1}{(1-\varepsilon^2)}\frac{\hbar^2k^2}{2m}$, to compensate the deviation, caused by the nonzero eccentricity. Hence, the time dependent widths maintain the initial ratio:
\begin{eqnarray}
	w_{b,t}^2=&\frac{1}{1-\varepsilon^2}\Bigg[w_{a}^2+\frac{t^2}{4w_{a}^2}\Bigg],\\
	w_{a,t}^2=&(1-\varepsilon^2) w_{b,t}^2.\label{eqDisp}
\end{eqnarray}
The above analysis is primarily meant for noninteracting case to pave a guideline for studying the interacting case. The influence of eccentricity on the quadratic dispersion, characterized by the term $\sqrt{1-\varepsilon^2}$,  in a non-interacting gas, also holds for interacting case. The time-evolved clouds in the interacting case will differ from that in noninteracting case. However, the width-ratio $\frac{w_{a,t}}{w_{b,t}}$ will remain unchanged, which suggests a similar dispersion management for interacting case too.\\
To mitigate the effects of nonzero eccentricity, it is necessary to manage dispersion such that the dispersion along the $y$-direction is slower than that along the $x$-direction by a factor of $(1-\varepsilon^2)$. This can be experimentally achieved by introducing a weak 2D optical lattice (OL) along with the elliptical waveguide \cite{eiermann2003dispersion,eiermann2004bright} with their lattice vectors satisfying $\frac{k_x}{k_y}=\sqrt{1-\varepsilon^2}$ \cite{kraemer2003bose,liang2008sound}. Moreover, the influence of ellipticity can be counteracted by adjusting the depth of the elliptical waveguide. The impact of varying width along the circumference can be nullified by modulating the waveguide's depth.

\section{Dispersion Management of Ground States}

After incorporating the appropriate dispersion management, the effective Gross-Pitaevskii equation becomes \cite{eiermann2003dispersion,eiermann2004bright,petrovic2010spatiotemporal,al2011analytical,yu2015matter,liu2017dark,liu2020analytical,atre2007controlling,sakaguchi2019interactions}:
\begin{eqnarray}
	i\frac{\partial \psi}{\partial t}=\Bigg[-\frac{\alpha}{2} \frac{\partial^2}{\partial x^2}-\frac{\beta}{2} \frac{\partial^2}{\partial y^2}+g|\psi|^2+V(x,y)\Bigg]\psi. \label{DMGPE}
\end{eqnarray}
Here, $\alpha$ and $\beta$ are the dispersion coefficients in $x$- and $y$- directions, respectively, whereas $V(x,y)$ is the potential of the elliptical waveguide.\\

\textit{\textbf{Numerical Method}}: The ground state solution of Eq.[\ref{DMGPE}] is numerically obtained by implementing the imaginary time propagation (ITP) method, where the initial wavefunction is allowed to evolve in imaginary time, $t=i\tau$. In this case, any initial wavefunction under the action of time evolution operator, $\exp{(-\tau \hat{H})}$, asymptotically converges to the ground state solution as $t\rightarrow \infty$ \cite{bader2013solving}. The time dynamics of a localized matter wave packet in the elliptical waveguide are obtained by the real time propagation (RTP) method. In both ITP and RTP methods, the linear and non-linear parts of the dynamical equation are treated separately, where the linear part is evolved in the momentum space, and the non-linear part is evolved in the coordinate space \cite{bao2006efficient}. The $x$ and $y$ coordinates are equally divided into 512 grids with a step size of 0.1. The step size for time is 0.08, totalling 16384 grids. In our work, we have considered ${}^{23}$Na BEC of $N=10000$ atoms, with parameters $m=3.816\times10^{-26}\;$kg, $\omega_{\perp}=512\;$Hz, $a_{\perp}=2.318\mu$m, and $a_{s}=2.75\times 10^{-9}$m \cite{wright2013driving,murray2013probing,jendrzejewski2014resistive}. The initial condensate in the form of binary peaks is placed diametrically opposite along the $x$-axis with the coordinates $(a,0)$ and $(-a,0)$, respectively.

\begin{figure*}[htpb]
	\centering
	\includegraphics[width=13 cm]{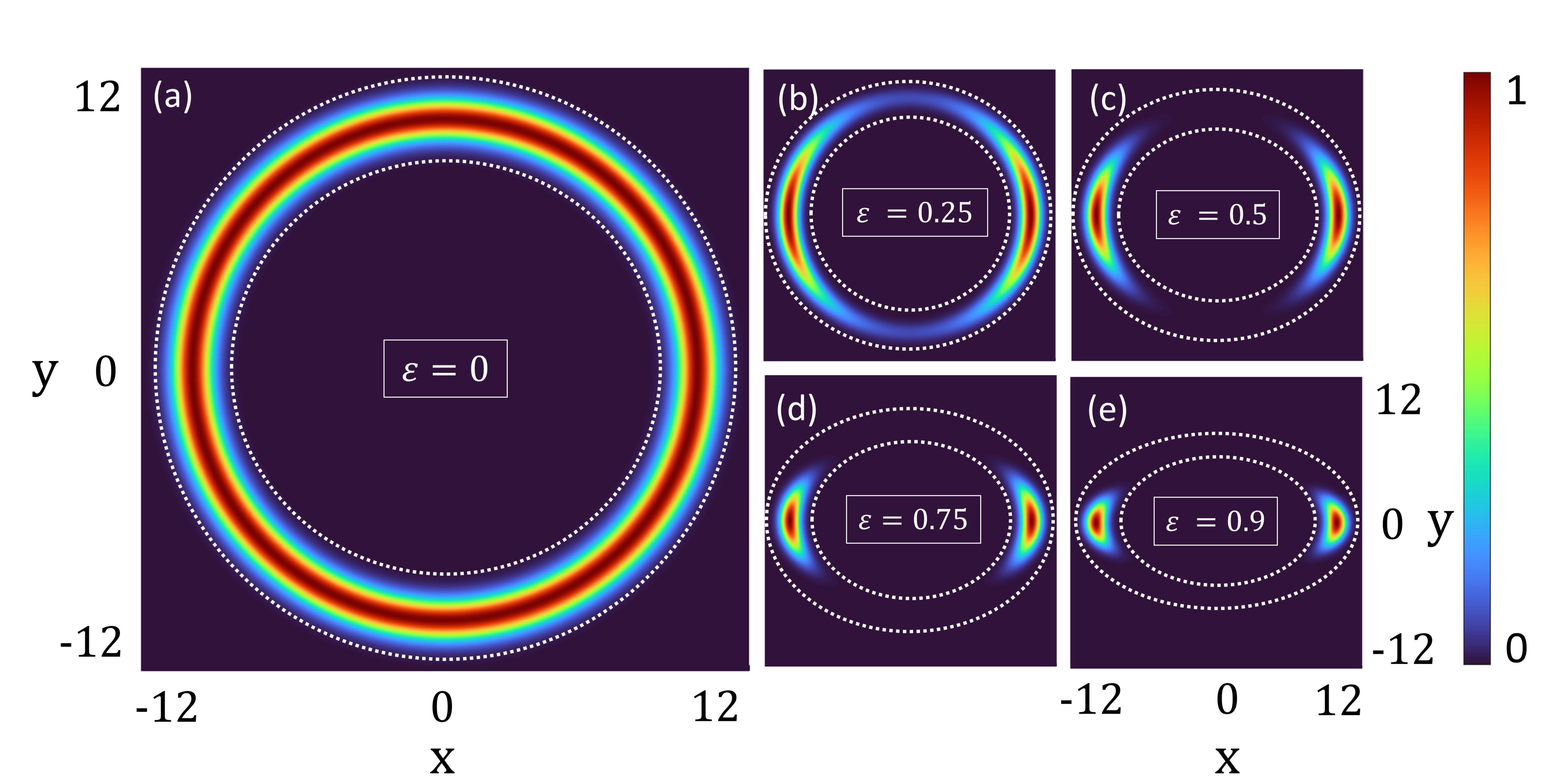}
	\caption{Ground State solution of BEC in (a) a circular waveguide, $\varepsilon=0$ and elliptical waveguides of eccentricity (b) $\varepsilon=0.25$, (c) $\varepsilon=0.5$, (d) $\varepsilon=0.75$ and (e) $\varepsilon=0.9$. The circular ring radius and the semimajor radius are taken as $a=10a_{\perp}$. $x$ and $y$ are in the units of $a_{\perp}=2.32\;\mu$m and, $t$ is in the units of $1/\omega_{\perp}=1.95\;$ms. The atomic density in each plot is scaled by its maximum and thus, the colorbar spans between [0,1].}
	\label{GS_sol}
\end{figure*}

\begin{figure*}[htpb]
	\centering
	\includegraphics[width=13. cm]{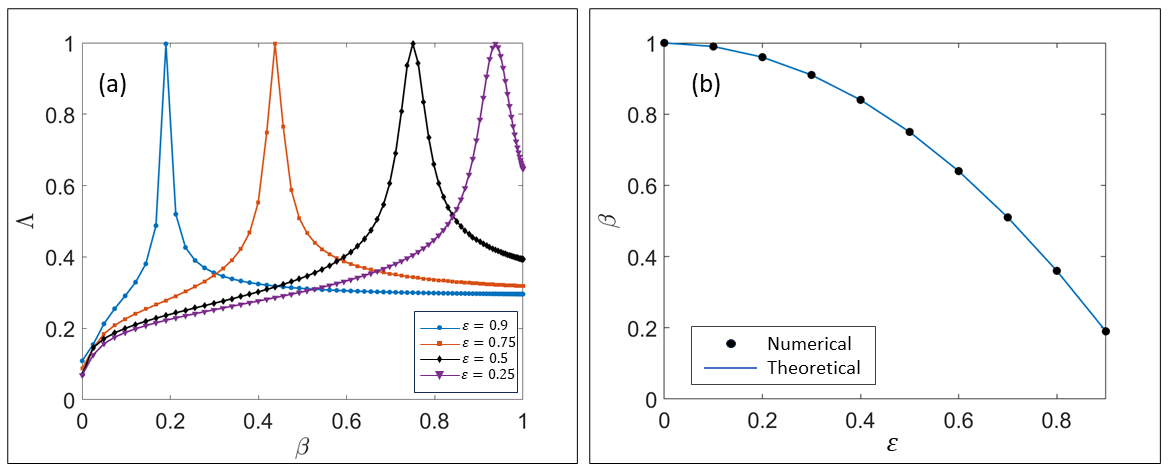}
	\caption{(a) Variation of overlap of the actual and desired wavefunction with the dispersion coefficient for various values of eccentricities. Solid lines with circles, squares, diamonds and triangles represent the eccentricities $\varepsilon=0.9$, $\varepsilon=0.75$, $\varepsilon=0.5$ and $\varepsilon=0.25$, respectively. (b) Numerical (dot) and theoretical (solid line) values of dispersion coefficient for various eccentricities. The circular ring radius and the semimajor radius are taken as $a=10a_{\perp}$.}
	\label{Overlap}
\end{figure*}

\subsection{Ground State of BEC in an Elliptical Waveguide}

For numerically finding the ground states, we consider a circular waveguide of radius $a=10a_{\perp}$ and unity dispersion coefficients ($\alpha=\beta=1$), for which the ground state density is uniform along the circumference of the ring. Figure \ref{GS_sol}(a) shows the ground state of the circular waveguide. The solution can be expressed as follows:
\begin{equation}
	\psi_{c}(x,y)=(\frac{\sqrt{V_0}}{\pi\gamma})^{\frac{1}{4}} e^{-\frac{\sqrt{V_0}(\sqrt{x^2+y^2}-a)^2}{2\gamma}}, \label{GS_circle}
\end{equation}
The ground state of the potential is a Gaussian ring since the cross-section of the potential in the vicinity of the minima is harmonic in nature. Such Gaussian ring condensate inside a circular waveguide has been discussed in various experimental and theoretical works \cite{kavoulakis2003bose,salasnich2022bose}. More interesting things happen when we increase the eccentricity of the ring waveguide from zero. The stationary states for waveguides with various eccentricities are shown in Fig.\ref{GS_sol}(b-e). The density is no longer uniformly distributed across the circumference of the waveguide, whereas one could see the density accumulation at the semi-major edges. The greater the eccentricity, the greater the density accumulation at the edges, thereby making the waveguide behave like a double-well potential. If the effects of ellipticity are counterbalanced, one could obtain a uniform stationary state in an elliptical waveguide.

\begin{figure*}[htpb]
	\centering
	\includegraphics[width=13 cm]{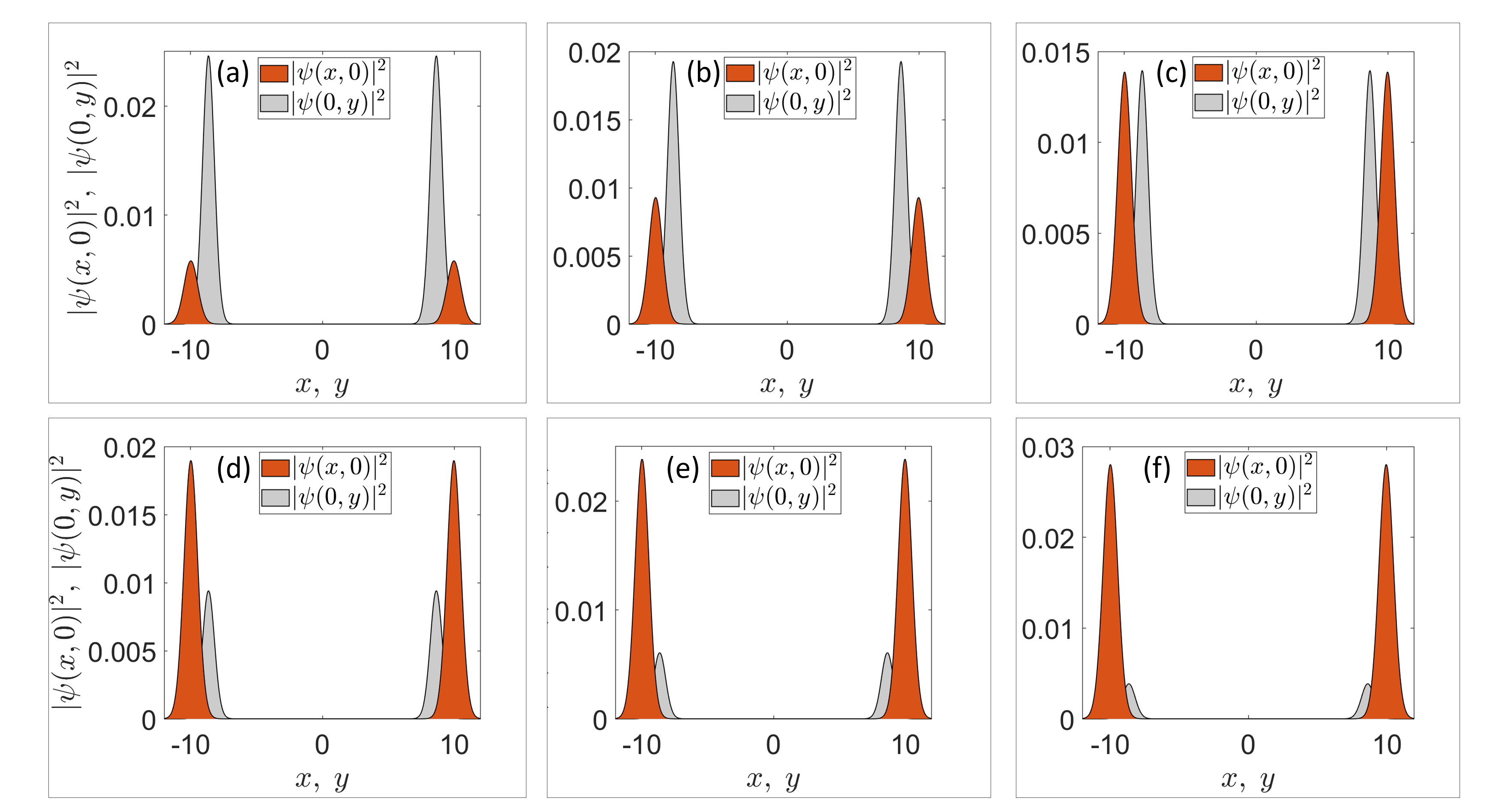}
	\caption{1D Cross-Sectional Densities in x and y directions for various $\beta$ values (a)$\;0.7250$, (b)$\;0.7375$, (c)$\;0.75=\beta_c$, (d)$\;0.7625$, (e)$\;0.7750$, (f)$\;0.7875$. The circular ring radius and the semimajor radius are taken as $a=10a_{\perp}$. $x$ and $y$ are in the units of $a_{\perp}=2.32\;\mu$m and, $t$ is in the units of $1/\omega_{\perp}=1.95\;$ms.}
	\label{CS_density}
\end{figure*}

\subsection{Coefficients of Dispersion to Obtain Uniform Stationary State}
As we increase the eccentricity of the waveguide, it gradually transforms to an effective double-well potential, resulting in a nonuniform stationary state along the circumference. As discussed earlier, to eliminate the effects of nonzero eccentricities, we employ the method of dispersion management. The dispersion coefficients $\alpha$ and $\beta$ are tuned, and here we keep $\alpha=1$ and vary $\beta$ to obtain a uniform stationary state in the elliptical waveguide. For the purpose of determining $\beta$, we find the overlap of the condensate density with an expected uniform density. The overlap between these two wavefunctions is defined by
\begin{equation}
	\Lambda=\frac{[\int_{-\infty}^{\infty}\int_{-\infty}^{\infty}|\psi_{e}(x,y)|^2|\psi_{a}(x,y)|^2 dx dy]^{2}}{\int_{-\infty}^{\infty}|\psi_{e}(x,y)|^4 dx dy \int_{-\infty}^{\infty}|\psi_{a}(x,y)|^4 dx dy}. \label{Overlap_eqn}
\end{equation}
Here, $\psi_e(x,y)$ is the expected wavefunction and $\psi_a(x,y)$ is the actual wavefunction obtained numerically. Here, the expected wavefunction is taken as elliptic Gaussian ring function following circular Gaussian ring for circular waveguide. Hence, the unity overlap function ($\Lambda=1$) will imply a uniform distribution of condensate density along the perimeter of the waveguide, whereas lower values of $\Lambda$ indicate deformations. Figure \ref{Overlap}(a) shows the variation of the overlap function with $\beta$ for different eccentricities. The maxima in the $\Lambda$ \textit{vs} $\beta$ curves will give us the necessary $\beta_c$ values to get the uniform ground state inside an elliptical waveguide. It is worth observing that $\beta_c$ becomes lower for higher eccentricities.

\begin{figure*}[htpb]
	\centering
	\includegraphics[width=13 cm]{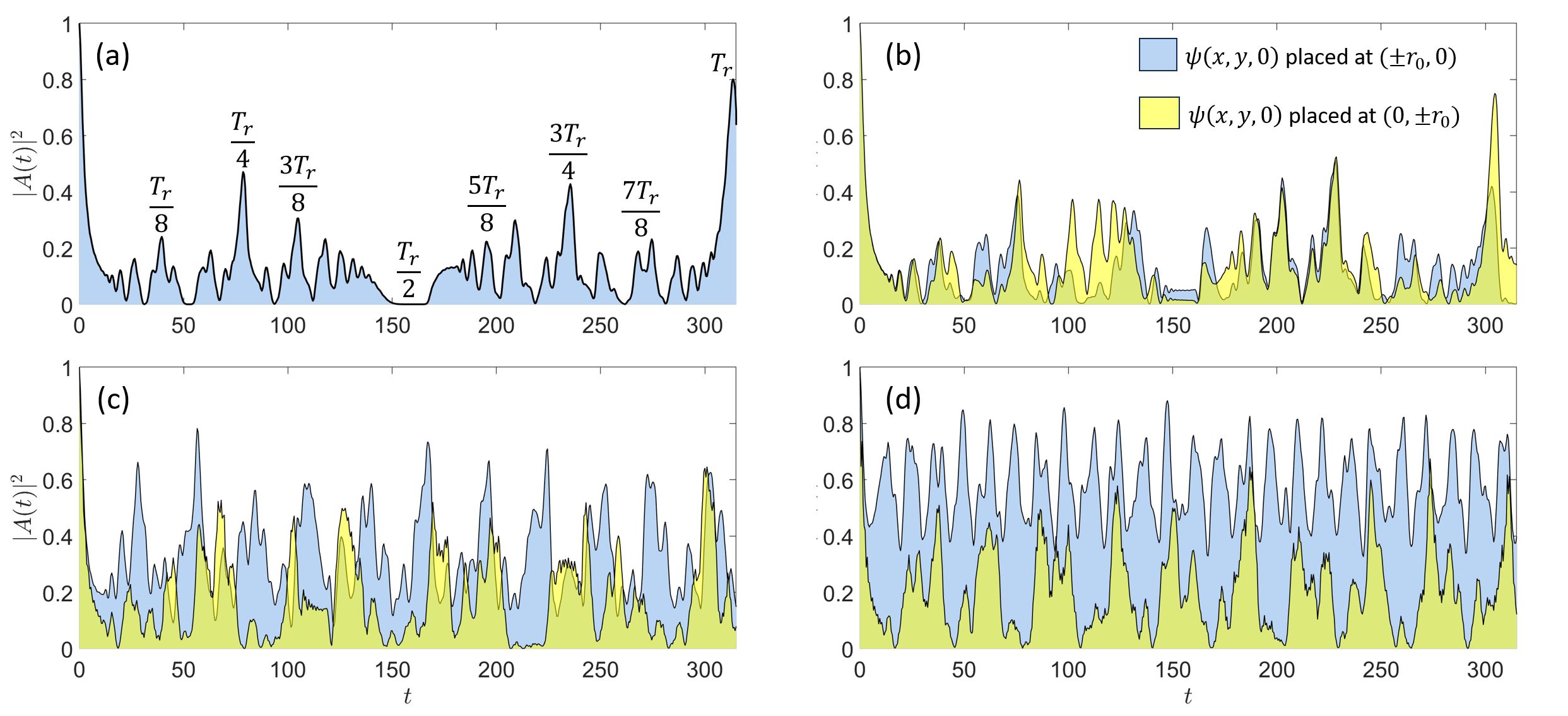}
	\caption{Survival function for two different initial orientations of the initial clouds $(\pm a,0)$ and $(0,\pm a)$, with interatomic interaction $g=2$ and, for eccentricities (a) $\varepsilon=0$, (b) $\varepsilon=0.25$, (c) $\varepsilon=0.75$ and (d) $\varepsilon=0.9$. The circular ring radius and the semimajor radius are taken as $a=10a_{\perp}$. $t$ is in the units of $1/\omega_{\perp}=1.95\;$ms.}
	\label{ACs}
\end{figure*}

\subsection{Dispersion Coefficients and Eccentricities}

We have noticed that for higher eccentricity, one needs to take lower $\beta_c$ to maintain the uniformity of the ground state. However, the exact relationship between $\beta_c$ and eccentricity is not obtained. To obtain this relation, we write Eq. \ref{DMGPE} with the transformed variable, $Y=\frac{y}{\sigma}$, where $\sigma=\sqrt{1-\varepsilon^2}$:
\begin{eqnarray}
	\Bigg[i\frac{\partial}{\partial t}+\frac{\alpha}{2} \frac{\partial^2}{\partial x^2}+\frac{\beta}{2\sigma^2} \frac{\partial^2}{\partial Y^2}-g|\psi|^2-V(x,Y)\Bigg]\psi=0, \nonumber
\end{eqnarray}
where $\psi\equiv\psi(x,Y)$. The potential of the elliptical waveguide transforms to
\begin{eqnarray}
	V(x,Y)=V_{0}\Bigg[1-e^{-\frac{1}{\gamma^2}(\sqrt{x^2+Y^2}-a)^2}\Bigg]. \label{Tpot}
\end{eqnarray}
Therefore, it becomes transparent to infer that, the dynamical equations for circular and elliptic cases take identical form provided
\begin{eqnarray}
	\frac{\beta_c}{\alpha_c}=1-\varepsilon^2 .\label{beta_cal}
\end{eqnarray}
This confirms the earlier prediction from the width dynamics of a Gaussian wavepacket in the vicinities of semi-major and semi-minor edges. The obtained $\beta_c$ is plotted along with it numerically obtained values in Fig.\ref{Overlap}(b), where the dots indicate the numerical values and the solid line indicates the values obtained from Eq.\ref{beta_cal}. It is clear that $\beta_c$ falls as we increase the eccentricity, such that $\beta_c=1$ for a circular waveguide and $\beta_c\rightarrow 0$ for higher eccentricities. The desired dispersion coefficient $\beta_c$, being a maximum in the overlap $\Lambda$, indicates that the ground state density below and above $\beta_c$ must be non-uniformly distributed and different from each other. This is visualized in Fig. \ref{CS_density} to find the cross-sectional densities in the elliptical ring along the semi-major and semi-minor axes. The 1D cross-sectional (CS) densities along the semi-major and semi-minor axes are denoted by $|\psi(x,0)|^2$ and $|\psi(0,y)|^2$, respectively. Figure \ref{CS_density} is depicted for an  eccentricity $\varepsilon=0.5$, where it is clear that at $\beta=\beta_c=0.75$, as shown in Fig.\ref{CS_density}(c), the 1D cross-sectional densities along the semi-major and semi-minor axis are almost equal, $|\psi(x,0)|^2\approx|\psi(0,y)|^2$. On the other hand, for $\beta<\beta_c$, $|\psi(x,0)|^2<|\psi(0,y)|^2$ (see Fig.\ref{CS_density}(a-b)) and for $\beta>\beta_c$, $|\psi(x,0)|^2>|\psi(0,y)|^2$ (see Fig.\ref{CS_density}(d-f)). Without any dispersion management, the dispersion coefficient is unity, $\beta=1$, corresponding to the complete density accumulation at the semi-major edges and $|\psi(0,y)|^2\approx0$. In the next section, we will study the implications of dispersion management by exploring the impact of waveguide eccentricity on the revival dynamics of matter waves.

\section{Fractional Revivals in an Elliptical Waveguide and Dispersion Management}

When a localised cloud of BEC is placed in a circular waveguide, it disperses and interferes with itself, forming interference fringes. The time at which the interference brings out the revival of the dispersed cloud in shape and position is termed the revival time $T_r$. At fractional multiples of the revival time, we can have multiple replicas of the initial condensate. This phenomenon is called fractional revivals. In this work, the initial condensate in the form of binary peaks is placed at $(\pm a,0)$. From the physics of dispersion, the revival time and FR time scales of two clouds in a circular waveguide of circumference, $C=2\pi r_0$, are given by \cite{bera2020matter}
\begin{eqnarray}
	T_{r}=\frac{C^2}{4\pi},\;\;\;
	t=\frac{p}{q}T_{r},\label{TrC}
\end{eqnarray}
where $p$ and $q$ are mutually prime integers. However, things are different when one places the binary peaks of BEC inside an elliptical waveguide of high eccentricity.

\begin{figure*}[htpb]
	\centering
	\includegraphics[width=13 cm]{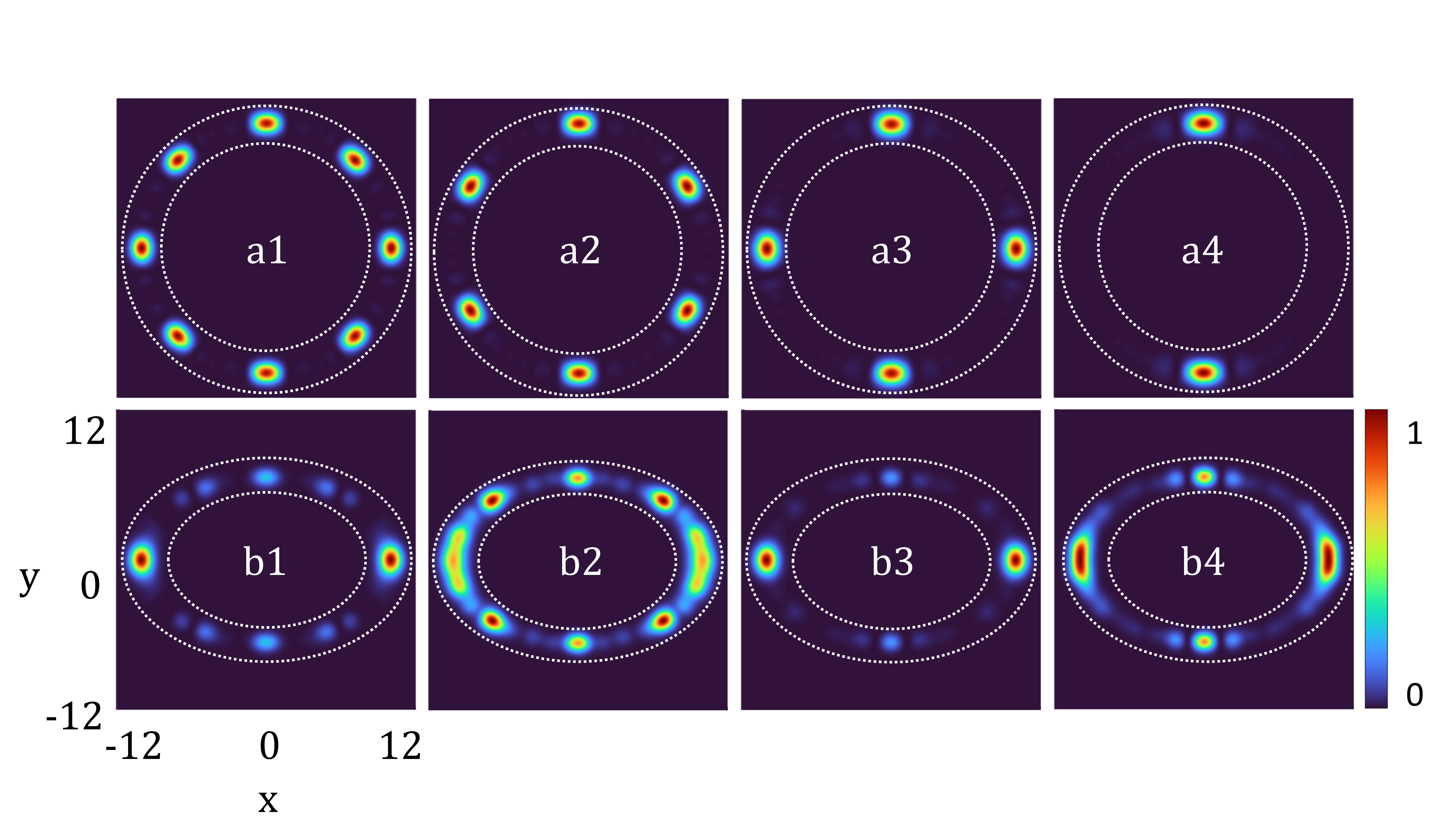}
	\caption{Condensate density in circular waveguide $\varepsilon=0$ at time instances $20.14\;$ms, $26.85\;$ms, $40.28\;$ms and $80.57\;$ms shown in a1, a2, a3 and a4, respectively. Condensate density in an elliptical waveguide of eccentricity $\varepsilon=0.75$ at time instances $14.50\;$ms, $19.30\;$ms, $28.95\;$ms and $57.90\;$ms shown in b1, b2, b3 and b4, respectively. The circular ring radius and the semimajor radius are taken as $a=10a_{\perp}$. $x$ and $y$ are in the units of $a_{\perp}=2.32\;\mu$m and, $t$ is in the units of $1/\omega_{\perp}=1.95\;$ms. The atomic density in each plot is scaled by its maximum and thus, the colorbar spans between [0,1].}
	\label{FR_density}
\end{figure*}

\subsection{Revival Time Scale for an Elliptical Waveguide}
The revival dynamics are conventionally studied by the time-dependent characteristic functions such as the autocorrelation function $A(t)$ or the survival function $S(t)$ \cite{eryomin1994manifestations,robinett2004quantum,ghosh2007time}. The survival function is the probability of finding the condensate in its initial state. In other words, it is the absolute square of the Autocorrelation function $A(t)$, which is defined as follows:
\begin{eqnarray}
	S(t)=|A(t)|^2,\\
	A(t)= {\int_{-\infty}^{\infty}\int_{-\infty}^{\infty}\psi^{*}(x,y,0)\psi(x,y,t)dx dy}.
\end{eqnarray}
It is a time series that quantifies the overlap of the wavefunction at a later time with that of the initial wavefunction, where its value closer to $1$ indicates full revival, and the smaller peaks indicate FR instances. In a circular waveguide, at half revival $T_r/2$ and its odd integral multiples, $|A(t)|^2$ becomes zero since the clouds are at position $(0,\pm a)$, which is spatially orthogonal to the initial location $(\pm a,0)$. Figure \ref{ACs} shows the survival function $|A(t)|^2$ for BEC in an elliptical waveguide of different eccentricities $(a)\;\varepsilon=0$, $(b)\;\varepsilon=0.25$, $(c)\;\varepsilon=0.75$, and $(d)\;\varepsilon=0.9$. Here, one could note that for $\varepsilon=0$, there is a clear signature of FR, as pointed out in Fig. \ref{ACs} (a). However, the signature of FR disappears at higher eccentricities. At eccentricity as high as $\varepsilon=0.9$, one could see many peaks with similar heights and the minimum of the survival function no longer touches zero, indicating that the cloud spends most of its time at the semi-major edges $(\pm a,0)$. This is delineated by showing the survival function for two different initial orientations of the clouds, namely $(\pm a,0)$ and $(0,\pm a)$, in Fig.\ref{ACs}.  At $\varepsilon=0$, the survival functions coincide for these two different orientations, whereas at higher eccentricities, the survival functions are no longer identical. While the survival function for $(0,\pm a)$ orientation touches zero more often, the survival function for $(\pm a,0)$ orientation hardly touches zero. Therefore, irrespective of where the initial clouds are placed inside the elliptical waveguide, the cloud tends to spend most of its time in the semi-major edges. This clearly indicates the disruption of FR instances in an elliptical waveguide. However, one would still get FR instances at very low eccentricities, and the corresponding time scales are worth finding out.

\subsection{Restoration of Fractional Revivals through Dispersion Management}
When the eccentricity of the waveguide is non-zero, the revival time takes the form,
\begin{equation}
	T_{r}=\frac{a^2[\int_{0}^{2\pi}\sqrt{1-\varepsilon^2\sin^2\phi} d\phi]^2}{4\pi},\label{TrE}
\end{equation}
since the circumference of an ellipse is given by,
\begin{eqnarray}
	C=a\int_{0}^{2\pi}\sqrt{1-\varepsilon^2\sin^2\phi} d\phi,
\end{eqnarray}
where $\phi\in [0,2\pi]$ is the azimuthal coordinate, and $a$, $b$ are semi-major and semi-minor radii, respectively.

\begin{figure*}[htpb]
	\centering
	\includegraphics[width=13 cm]{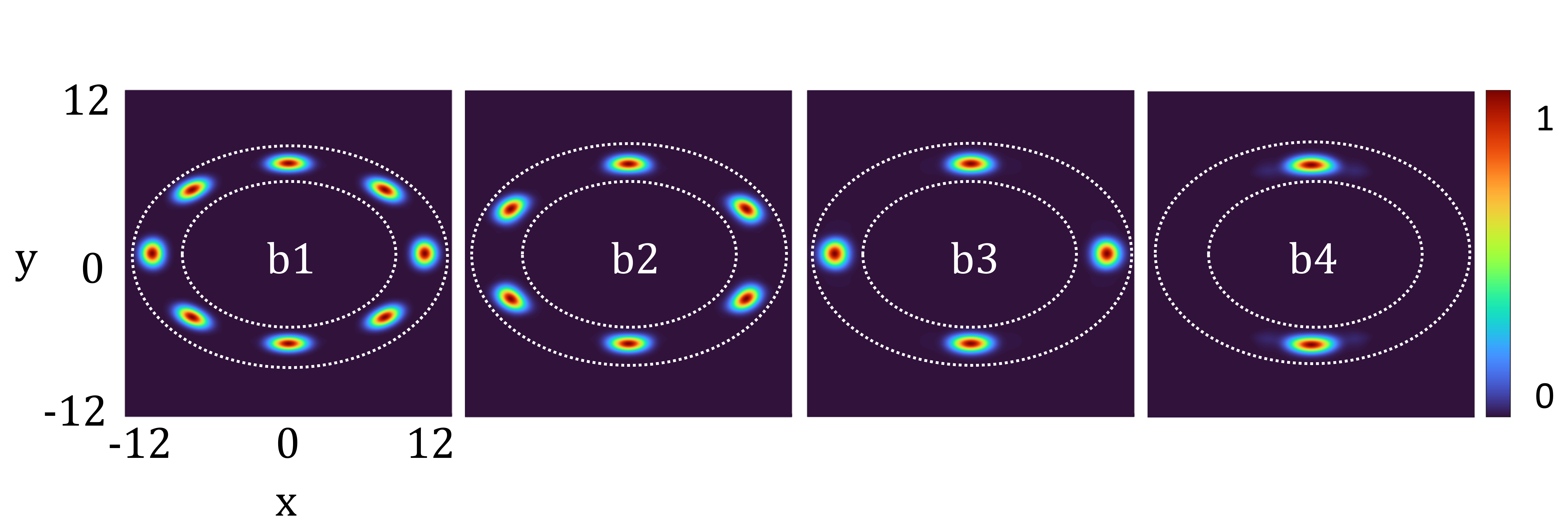}
	\caption{Snapshots of DM condensate density in elliptical waveguide of eccentricity $\varepsilon=0.75$ at time instances $20.14\;$ms, $26.85\;$ms, $40.28\;$ms and $80.57\;$ms shown in b1, b2, b3 and b4, respectively. The semimajor radius is $a=10a_{\perp}$. $x$ and $y$ are in the units of $a_{\perp}=2.32\;\mu$m and, $t$ is in the units of $1/\omega_{\perp}=1.95\;$ms. The atomic density in each plot is scaled by its maximum and thus, the colorbar spans between [0,1].}
	\label{FR_DM_0.75}
\end{figure*}
\begin{table*}[ht]
	\begin{tabular}{ | c | c | c | c | c | }
		\hline
		\boldmath \hspace{.5cm} Eccentricity \hspace{.4cm} & \hspace{.4cm} $\varepsilon=0$ \hspace{.3cm} & \hspace{.3cm} $\varepsilon=0.25$ \hspace{.3cm} & \hspace{.3cm} $\varepsilon=0.75$ \hspace{.35cm} & \hspace{.3cm} $\varepsilon=0.9$ \hspace{.3cm}\\ \hline
		$T_{r}$ before DM (ms) & $161.10$ & $156.07$ & $115.80$ & $95.86$ \\  \hline
		$T_{r}$ after DM (ms) & $161.10$ & $161.10$ & $161.10$ & $161.10$ \\ \hline
	\end{tabular}
	\caption{Revival time for different eccentricities before and after dispersion management. The circular ring radius and the semimajor radius are taken as $a=10a_{\perp}$.}
	\label{table1}
\end{table*}
At low eccentricities, the daughter condensates of the FR are spatially resolved, and the revival time can be defined by Eq.\ref{TrE}. However, at higher eccentricities, the multiple splits of the FR are no longer spatially resolved, and the FR patterns are disrupted. This is evident from Fig. \ref{FR_density}, where the condensate densities at times $\frac{T_{r}}{8}$, $\frac{T_{r}}{6}$, $\frac{T_{r}}{4}$, and $\frac{T_{r}}{2}$ are denoted by numbers $1,\;2,\;3,\;4$, respectively. Figures for the two distinct cases ($\varepsilon=0$ and $0.75$) are consequently leveled by $(a)$ and $(b)$. One can observe no-FR for $\varepsilon=0.75$ and the cloud tends to spend more time at the semi-major edges, irrespective of the initial placement of the cloud. In the previous section, we showed that choosing the appropriate dispersion coefficients could nullify the effects of the non-constant curvature and produce uniformly distributed ground states inside an elliptical waveguide. We apply this technique to restore fractional revivals of matter waves in an elliptical waveguide. We choose the dispersion coefficients $\alpha=1$ and $\beta=1-\varepsilon^2$, as obtained in Eq. \ref{beta_cal}. The spatial resolution of the daughter condensates at FR instances, for $\varepsilon=0.75$, is far lower than that in $\varepsilon=0.5$. In such a case, the significance of dispersion management is greatly evident. Figure \ref{FR_DM_0.75} shows the snapshots of dispersion managed condensate density for eccentricity $\varepsilon=0.75$ at FR instances $\frac{Tr}{8},\;\frac{Tr}{6},\;\frac{Tr}{4}$ and, $\frac{Tr}{2}$ denoted by $b1,\;b2,\;b3$ and, $b4$, respectively. The snapshots confirm the restoration of FR patterns of dispersion-managed BEC in an elliptical waveguide. The daughter condensates at the FR instances are spatially resolved in the case of dispersion-managed BEC. Interestingly, for a dispersion-managed matter wave, the revival time and FR times no longer depend on the eccentricity, unlike the non-dispersion-managed case. Table \ref{table1} shows the revival times for different eccentricities without and with dispersion management (DM). One could note that after DM, the BEC revives at times independent of the eccentricity of the waveguide. The eccentricity-dependent dispersion coefficient (Eq.\ref{beta_cal}) balances the eccentricity-dependent time scale (Eq.\ref{TrE}) of the matter wave. In other words, the ellipticity-induced effects are nullified through dispersion management, and the matter-wave in the elliptical waveguide of semi-major radius $a$ behaves like that in a circular waveguide of radius $a$. Next, we will focus on atom interferometry, specifically investigating how these FR clouds, restored through dispersion management, can be utilized for quantum sensing.

\begin{figure*}[htpb]
	\centering
	\includegraphics[width=13 cm]{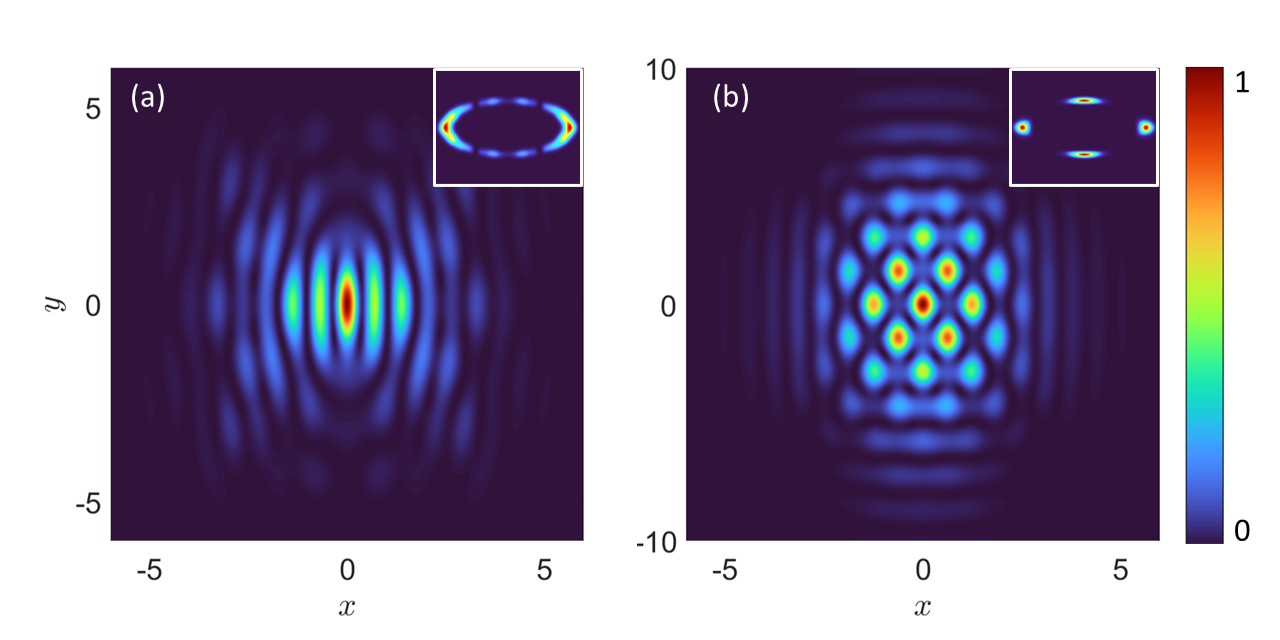}
	\caption{The interference patterns, obtained by deactivating the elliptical waveguide of eccentricity, $\varepsilon=0.9$ at $t=\frac{T_r}{4}$. (a) Without dispersion management, $\alpha=\beta=1$ (b) After dispersion management, $\alpha=1$, $\beta=1-\varepsilon^2$. The inset shows the corresponding condensate densities. The semimajor radius is taken as $a=10a_{\perp}$. $x$ and $y$ are in the units of $a_{\perp}=2.32\;\mu$m and, $t$ is in the units of $1/\omega_{\perp}=1.95\;$ms. The atomic density in each plot is scaled by its maximum and thus, the colorbar spans between [0,1].}
	\label{IP_1}
\end{figure*}

\section{Atom Interferometry in Elliptical Waveguide}

The restored fractional revivals in an elliptical ring can have significant applications in quantum sensing, where the FR of the initially localized matter wave acts as a matter wave beam splitter. Specifically, at the quarter revival time, a localized initial double peak BEC splits into four daughter condensates. The phase information can be read from the interference pattern obtained by switching off the waveguide and allowing the condensate to interfere in a harmonic trap of frequency, $\omega_{ho}=0.5\omega_{\perp}$. The harmonic trap is used to obtain the interference pattern within a confined region. The interference pattern is captured at the evolution time of $t_{f}=5.85\;$ms. In this work, we obtain the interference patterns for 4 split FR clouds in an elliptical waveguide of eccentricity, $\varepsilon=0.9$.

In elliptical waveguides of eccentricity as high as $\varepsilon=0.9$, an initially localized BEC does not undergo a quarter revival split into four daughter condensates, as shown in the inset of Fig.\ref{IP_1}(a). In such a case, the interference pattern obtained does not show any known 2D structures. However, through dispersion management $\beta=\beta_c$, one could restore the 4 splits in an elliptical waveguide (see the inset Fig.\ref{IP_1}(b)) and thereby obtain the rectangular lattice structured interference pattern. Figure \ref{IP_1}(b) shows the interference pattern for 4 split condensates with lattice periods $u=2.85\;\mu$m and $v=6.55\;\mu$m along $x$ and $y$ directions, respectively. It should be mentioned that the ratio of the lattice periods of the interference pattern is equal to the ratio of the semi-major and semi-minor radius, i.e., $\frac{v}{u}=\frac{b}{a}$. Since we know that the ratio of the semi-major and semi-minor radius is $\frac{b}{a}=\sqrt{1-\varepsilon^2}$, the lattice periods are related to the eccentricity of the elliptical waveguide as $\frac{v}{u}=\sqrt{1-\varepsilon^2}$. The interferometric fringes can be made further precise by increasing the radii of the ellipse and also the order of fractional revivals.
\\

\section{Conclusion}
We investigated the influence of ellipticity on the ground state of a BEC within an elliptical waveguide and its impact on the FR instances in a localized matter wave. The elliptical waveguide exhibits behavior reminiscent of a double-well potential. Notably, an increase in eccentricity correlates with a higher concentration of condensate density at these semimajor edges. We effectively manage dispersion to counteract the effects of variable thickness within the elliptical waveguide. We achieve a uniform ground state by identifying optimal dispersion coefficients from the overlap function. Interestingly, these coefficients are found to be contingent upon the waveguide's eccentricity. There are several schemes by which such dispersion management can be realized for atomic BECs \cite{louis2005dispersion,barontini2007dynamical,khamehchi2017negative,colas2018negative}. We have explored the disruption of FR patterns in a localized matter wave confined within an elliptical waveguide. We employ the time-dependent characteristic function known as the survival function to analyze the perturbed dynamics of the FR instances. Further, using the determined dispersion coefficients, we restore the FR instances of the Bose-Einstein condensate within the elliptical waveguide. Finally, the restored FR clouds are subjected to evolution in a harmonic potential, and their interference patterns are examined. This comprehensive study sheds light on the intricate interplay between ellipticity, dispersion and interferometry of matter waves and offers exchanging the physical merits between the atomtronics applications with zero and nonzero eccentricities.

\section{Acknowledgement}
We acknowledge Luca Salasnich, University of Padova for useful discussions. UR acknowledges the support through the project (CRG/2022/007467) by the Science and Engineering Research Board, India.

\bibliographystyle{elsarticle-num}

\bibliography{ref}


\end{document}